\documentclass[aps,reprint,prl,amsmath,amssymb,floatfix,noeprint,superscriptaddress]{revtex4-2}

\usepackage{graphics}
\usepackage{natbib}
\usepackage{graphicx}
\usepackage{epsfig}
\usepackage{amsmath}
\usepackage{bm}
\usepackage{lineno}
\usepackage{color}
\usepackage{dcolumn}
\begin{document}
\title{Anomalous temperature dependence of phonon pumping by ferromagnetic resonance in Co/Pd multilayers with perpendicular anisotropy}
\author{W.~K.~Peria}
\affiliation{School of Physics and Astronomy, University of Minnesota, Minneapolis, Minnesota 55455, USA}
\author{D.-L.~Zhang}
\affiliation{Department of Electrical and Computer Engineering, University of Minnesota, Minneapolis, Minnesota 55455, USA}
\author{Y.~Fan}
\affiliation{Department of Electrical and Computer Engineering, University of Minnesota, Minneapolis, Minnesota 55455, USA}
\author{J.-P.~Wang}
\affiliation{Department of Electrical and Computer Engineering, University of Minnesota, Minneapolis, Minnesota 55455, USA}
\author{P.~A.~Crowell}
\affiliation{School of Physics and Astronomy, University of Minnesota, Minneapolis, Minnesota 55455, USA}
\begin{abstract}
We demonstrate the pumping of phonons by ferromagnetic resonance in a series of [Co(0.8~nm)/Pd(1.5~nm)]$_n$ multilayers ($n =$~6, 11, 15, and 20) with large magnetostriction and perpendicular magnetic anisotropy. The effect is shown using broadband ferromagnetic resonance over a range of temperatures (10 to 300~K), where a resonant damping enhancement is observed at frequencies corresponding to standing wave phonons across the multilayer. The strength of this effect is enhanced by approximately a factor of 4 at 10~K compared to room temperature, which is anomalous in the sense that the temperature dependence of the magnetostriction predicts an enhancement that is less than a factor of 2. Lastly, we demonstrate that the damping enhancement is correlated with a shift in the ferromagnetic resonance field as predicted quantitatively from linear response theory.
\end{abstract}
\maketitle
The ability to couple the spin degree of freedom with other degrees of freedom, such as charge or strain, is crucial to many spintronic applications. The coupling of spin to strain is a phenomenon known as magnetostriction, which is known to directly influence magnetization dynamics \cite{Bommel1959b,Seavey1965,Weber1968,Jager2013,Jager2015,Holanda2018,An2020,Zhao2020,Zhang2020d,Peria2021a}. Some work on dynamical magnon-phonon coupling has focused on the generation of phonons by ferromagnetic resonance (FMR) in a magnetic thin film and subsequent propagation of the phonons into the substrate, which is referred to as phonon pumping \cite{Seavey1965,Streib2018,Ruckriegel2020,Zhang2020b,Rezende2021}. Much of the early work on phonon pumping lacked broadband frequency dependence, which is necessary for fully characterizing the effect as well as demonstrating the existence of multiple resonances. Recent experimental work on phonon pumping has largely relied on time-resolved Kerr measurements \cite{Jager2013,Jager2015,Zhang2020d}, which are susceptible to strain excitation through laser heating rather than due to magnetization dynamics alone. Also, the temperature dependence of this effect has not been studied, which may provide new insights into the underlying physics.

In this Letter, we demonstrate the phonon pumping effect by ferromagnetic resonance in a series of [Co/Pd]$_n$ multilayers with perpendicular magnetic anisotropy (PMA). It is shown that the strength of the effect is strongly temperature dependent (a factor of $\sim$ 4 enhancement at 10~K relative to 300~K)---much more than would be expected from the temperature dependence of the magnetostriction alone (less than a factor of 2 enhancement)---which we argue is due to the sensitivity of the phonon pumping to the pinning of the dynamic magnetization. We also show that the frequencies of the phonon pumping resonances can be tuned by varying $n$, the number of Co/Pd repetitions. Finally, we show the dispersive effect of the phonon pumping through shifts in the FMR field, as predicted from the dissipation using Kramers-Kronig relations.

\begin{figure}
  \centering
  \includegraphics{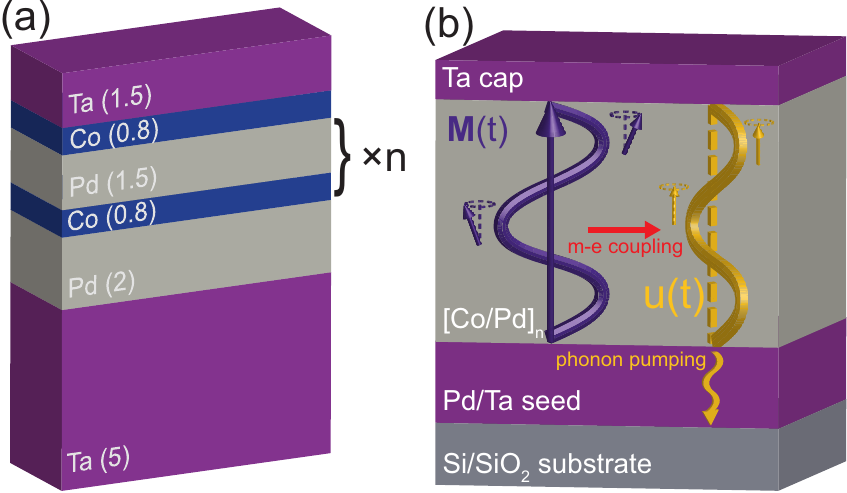}
  \caption{(a) Stack structure of the [Co/Pd]$_n$ multilayers. Thicknesses of each layer are given in parentheses and have units of nm. The Co(0.8~nm)/Pd(1.5~nm) bilayer is repeated a total of $n$ times as indicated on the figure. (b) Schematic of the phonon pumping process in the configuration where the magnetization $\mathbf{M}(t)$ is normal to the plane of the film. The magnetization depth profile is given by a sine wave (for simplicity) with pinning at the interfaces. The magnetoelastic coupling (shown by the red arrow) leads to the creation of a phonon standing wave with displacement $u(t)$. The phonon pumping process is shown by the wavy gold arrow representing the leakage of phonons into the seed layers and substrate.}\label{fig:fig1}
\end{figure}

Co/Pd multilayers are well-known for their large magnetostriction and PMA \cite{Hashimoto1989} and have been demonstrated for use in perpendicular magnetic tunnel junctions (p-MTJ) \cite{Tadisina2010}, including cases where synthetic antiferromagnets (SAF) made from Co/Pd multilayers were used for the reference layers \cite{Natarajarathinam2012,Chang2013}. The PMA is particularly significant for this application since phonon pumping is more efficient when the magnetization is perpendicular to the film plane \cite{Streib2018}. [Co(0.8~nm)/Pd(1.5~nm)]$_n$ multilayers ($n =$~6, 11, 15, and 20) were grown by dc magnetron sputtering at room temperature with a base pressure of $<5\times10^{-8}$~Torr using Ar gas at a working pressure of 2.0~mTorr. The thicknesses of the Co and Pd layers are 0.8~nm and 1.5~nm, respectively, for all of the samples and will henceforth be omitted. Ferromagnetic resonance of the [Co/Pd]$_n$ multilayers [Fig.\ \ref{fig:fig1}(a)] was measured using a coplanar waveguide setup with modulation of the applied magnetic field for lock-in detection of the transmitted microwave power, which was rectified with a Schottkey diode detector. The applied magnetic field was swept with the microwave frequency held fixed. Magnetometry measurements were performed on all the multilayers using superconducting quantum interference device (SQUID) magnetometry. The SQUID measurements were performed over a range of temperatures (5 to 300~K) for both in-plane and out-of-plane applied fields.  They confirmed an out-of-plane easy axis in all the samples, and was also used to measure the saturation magnetization as a function of temperature in the multilayers \cite{Supplemental}.

\begin{figure}
  \includegraphics{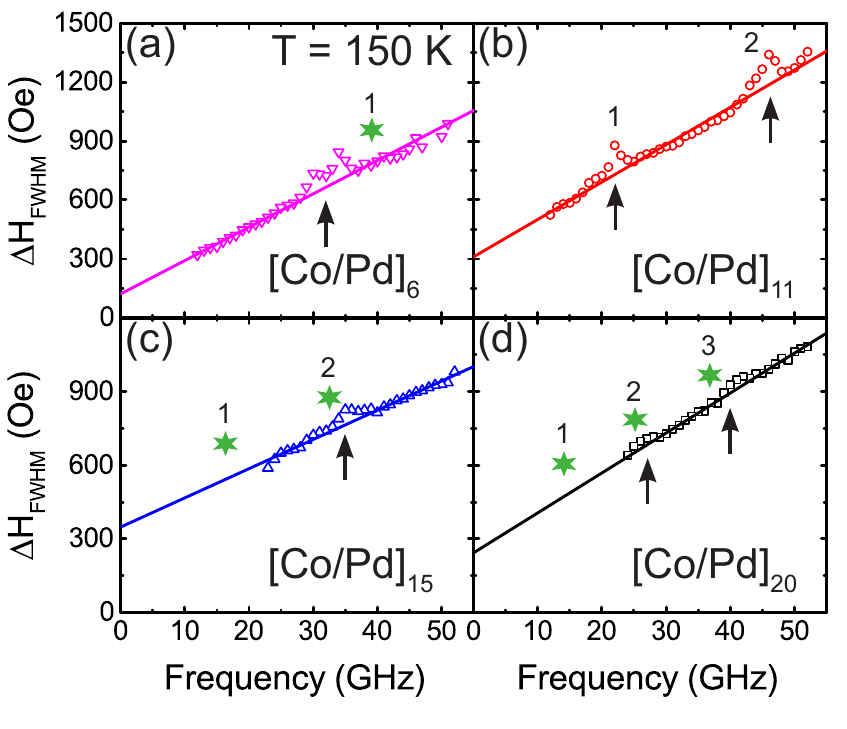}
  \caption{Ferromagnetic resonance linewidths as a function of frequency with applied magnetic field out of plane at $T = 150$~K for (a) [Co/Pd]$_6$ (magenta triangles), (b) [Co/Pd]$_{11}$ (red circles), (c) [Co/Pd]$_{15}$ (blue triangles), and (d) [Co/Pd]$_{20}$ (black squares). The vertical arrows indicate the positions of the phonon pumping resonances, and the green stars indicate the corresponding positions predicted from the positions observed in the [Co/Pd]$_{11}$ multilayer. The numbers labelling the stars correspond to the number of half-waves in the thickness resonance so that, e.g., ``3" means a phonon standing wave with wavelength $\lambda = 3d/2$, where $d$ is the thickness of the magnetic portion of the multilayer.}\label{fig:150Klinewidths}
\end{figure}

We first demonstrate the effect of phonon pumping on the FMR linewidths and how it depends on the number of Co/Pd repetitions in the multilayer stack. Fig.\ \ref{fig:fig1}(b) shows a schematic of the phonon pumping process, where magnetization dynamics are damped by the leakage of magnetoelastically-driven phonons into the substrate. Figure \ref{fig:150Klinewidths} shows FMR linewidths measured in a perpendicular field as a function of frequency at 150~K for four different [Co/Pd]$_n$ multilayer structures with $n =$~6, 11, 15, and 20. The lower frequency limit of the measurements is determined by the perpendicular anisotropy field (which sets the zero-field FMR frequency) for the $n = $~6 and 11 samples. For the $n = $~15 and 20 samples, the FMR signal disappears at nonzero field, which suggests that the sample becomes unsaturated at fields higher than zero. This observation is corroborated by out-of-plane magnetic hysteresis loops, which show the nucleation of domains before zero field is reached \cite{Supplemental}.

For all of the samples shown in Fig.\ \ref{fig:150Klinewidths}, there are resonant linewidth enhancements that appear at specific frequencies. The linear background is due to the Gilbert damping, for which fits were generated by excluding points within 3~GHz of the center of the peaks. For the $n=$~11 and 20 multilayers, there are two resonant peaks in the linewidth. In the $n=$~11 multilayer, the frequency of the high frequency peak is double that of the low frequency peak, implying that these represent the first and second harmonics of a fundamental damping resonance. In the $n=$~20 multilayer, the high frequency peak is  3/2 that of the low frequency peak, suggesting that the low and high frequency peaks are the second and third harmonics of a fundamental damping resonance, respectively. We cannot observe the fundamental resonance, however, since it is expected at a frequency ($\simeq$13 to 14~GHz) at which the sample is unsaturated. For the $n=$~6 and 15 multilayers, there is one peak in the linewidth. This corresponds to the fundamental damping resonance in the $n=$~6 multilayer, and the second harmonic in the $n=$~15 multilayer. The fundamental resonance is undetectable in the $n=$~15 multilayer because it occurs at a frequency ($\simeq$16~GHz) at which the sample is unsaturated. We note that the damping resonance in the [Co/Pd]$_6$ multilayer exhibits a twin-peak structure, with the two peaks separated by approximately 4~GHz. This may be due to the existence of standing waves with nodes at both the interface between the 2-nm Pd and 5-nm Ta seed layers \textit{and} the inteface between the Co and 2-nm Pd seed layer [shown in Fig.\ \ref{fig:fig1}(a)]. Were this the case, one would expect a spacing of about 4~GHz as we observe. This hypothesis predicts a peak spacing of $\lesssim$1~GHz for the thicker multilayers, which would explain why the twin-peak structure is only observed in the [Co/Pd]$_6$ multilayer.

The vertical arrows in Fig.\ \ref{fig:150Klinewidths} indicate the positions of the resonances for each multilayer. Transverse acoustic phonon standing waves are expected at frequencies where $d$, the thickness of the stack excluding capping and seed layers, matches an integer number of phonon half wavelengths. This condition can be expressed as $f = c_t/(2d/m)$ ($c_t$ is the transverse speed of sound and $m$ is a positive integer). Longitudinal phonons are neglected because they couple to the magnetization at higher order \cite{Kittel1958,Streib2018,Sato2021a}. The hypothesis that the multilayer is a half-wave resonator is based on the fact that the highly dense Ta capping and seed layers will lead to pinning of the phonons at these interfaces. The green stars in panels (a), (c), and (d) indicate the positions of the resonances predicted from the positions observed in the [Co/Pd]$_{11}$ multilayer in panel (b), where the effect is strongest. The numbers labelling the stars indicate the order of the resonance, so that a resonance of order $m$ corresponds to a phonon standing wave of wavelength $\lambda = 2d / m$. From this we note that there is good agreement between the observed and predicted positions of the resonances, which demonstrates that the damping resonances can indeed be thought of as ``thickness'' resonances. The most significant deviation is observed in the [CoPd]$_6$ sample, which is the thinnest and therefore most sensitive to changes in the effective thickness at the top and bottom interfaces.  We did not observe any thickness resonances for IP magnetization (shown in the Supplemental Material for the [Co/Pd]$_{11}$ multilayer \cite{Supplemental}), which is expected due to the fact that the strongest coupling is to phonons propagating parallel to the static magnetization \cite{Kittel1958,Streib2018,Sato2021a}.

\begin{figure}
  \centering
  \includegraphics{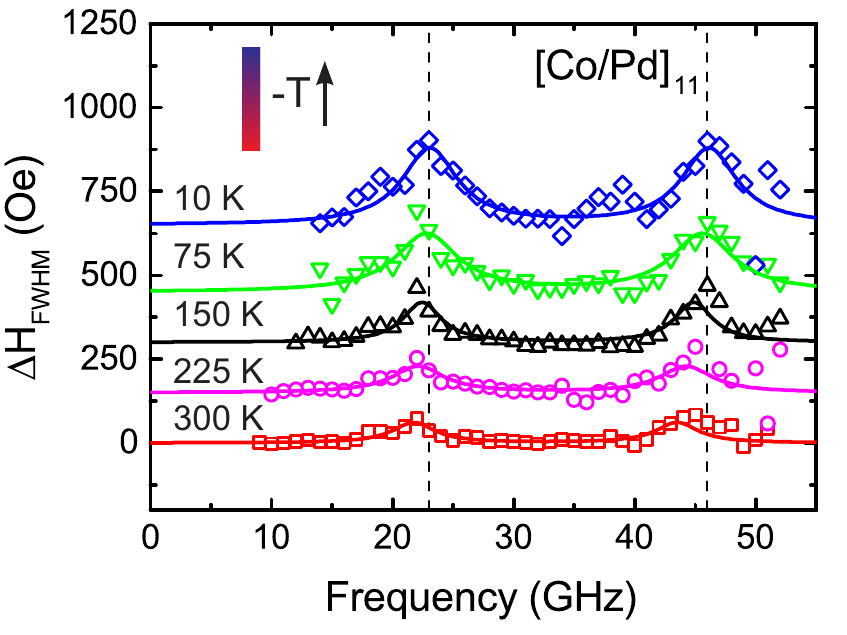}
  \caption{Evolution of the phonon pumping contribution to the FMR linewidths with temperature for the [Co/Pd]$_{11}$ multilayer at temperatures of 10~K (blue diamonds), 75~K (green triangles), 150~K (black triangles), 225~K (magenta circles), and 300~K (red squares). The vertical dashed lines indicate the locations of the phonon pumping peaks, 23~GHz and 46~GHz (at 10~K), which correspond to phonon wavelengths of $\lambda = 2d$ and $\lambda = d$, respectively, where $d$ is the thickness of the multilayer (excluding capping and seed layers). The data below 300~K are offset vertically so that the individual datasets could be more easily distinguished.}\label{fig:CP11alltemps}
\end{figure}

The temperature dependence of the phonon pumping contribution to the FMR linewidths of the [Co/Pd]$_{11}$ multilayer is shown in Fig.\ \ref{fig:CP11alltemps} for temperatures ranging from 10 to 300~K. The phonon pumping contribution is quantified by fitting the full width at half maximum (FWHM) FMR linewidths to the form
\begin{equation}\label{eq:pplinewidth}
  \Delta H_{FWHM} = \Delta H_0 + 2 \alpha \omega / \gamma + \Delta H_{ph}(\omega)~,
\end{equation}
where $\Delta H_0$ is the frequency-independent inhomogeneous broadening, $2 \alpha \omega / \gamma$ is the contribution from Gilbert damping ($\alpha$ is the Gilbert damping constant and $\gamma$ is the gyromagnetic ratio), and $\Delta H_{ph}(\omega)$ is the nonlinear frequency-dependent contribution from phonon pumping. We assume a phenomenological Lorentzian lineshape for the form of $\Delta H_{ph}(\omega)$:
\begin{equation}\label{eq:absorptive}
  \Delta H_{ph} (\omega) = \sum_n 2A_n \frac{\delta \omega / 2}{(\omega-n\omega_0)^2 + (\delta \omega / 2)^2}~,
\end{equation}
where $\delta \omega$ is the FWHM of the resonance, $\omega_0$ is the frequency of the fundamental half-wave resonance, and $A_n$ sets the amplitude (the factor of 2 is needed to convert from HWHM to FWHM). In the case of the [Co/Pd]$_{11}$ multilayer, we enforce the constraints that the high frequency resonance is exactly twice the low frequency resonance and that the amplitudes of both resonances are equal. The widths of the resonances are set by the acoustic impedance ratios at the boundaries, so that a strong mismatch will yield a sharp resonance \cite{Comstock1963,Seavey1965,Weber1968,Seavey1968,Streib2018}, and should not depend on the order of the resonance. The phonon relaxation rate can also influence the resonance width \cite{Comstock1963,Weber1968,Seavey1968}, but this is probably a secondary effect \footnote{The phonon mean free path is likely longer than the multilayer thicknesses ($\sim$10 to 40~nm), so the effect of phonon relaxation within the multilayer probably has a negligible effect on the phonon pumping resonances. The widths of the resonances are also largely independent of temperature, which is untrue of the phonon relaxation.}.

It is clear from Fig.\ \ref{fig:CP11alltemps} that the intensity of the thickness resonances increases strongly at low temperature (whereas the Gilbert damping depends very weakly on temperature, as shown in the Supplemental Material \cite{Supplemental}). The amplitudes of the resonances are about a factor of 4 larger at 10~K relative to 300~K. Magnetoelastic effects, originating from a magnetic anisotropy energy, are expected to increase at low temperature due to a reduction of thermal fluctuations of the magnetization \cite{Callen1966,Peria2021a}. A strong temperature dependence of the amplitude of the resonances is seen for all of the multilayers, increasing at low temperature by a magnitude similar to that seen in Fig.\ \ref{fig:CP11alltemps} for the [Co/Pd]$_{11}$ sample. Also noteworthy is the small upward shift in the frequency of the resonances with decreasing temperature. This is consistent with the expectation that the elastic moduli should increase at low temperature, causing an increase in the speed of sound (which is proportional to the frequency of a given thickness resonance). The frequencies of the first and second thickness resonances shift from 22 and 44~GHz to 23 and 46~GHz, respectively.

There has been significant theoretical work attempting to model phonon pumping \cite{Seavey1963,Comstock1963,Kooi1963,Seavey1968,Kobayashi1973a,Streib2018,Zhang2020b,Sato2021a}, and so we will not give a comprehensive overview here. All models predict that the phonon pumping amplitude should go as the square of the magnetoelastic coefficient, which can be understood in terms of Fermi's golden rule. One of the primary factors influencing the phonon pumping is the nonuniformity of the dynamic magnetization, which is necessary for exciting acoustic phonons (since they have nonzero wave vector). The model presented by \citet{Streib2018} assumes uniform magnetization within the film, with the only nonuniformity coming from the discontinuity of the magnetization at the interfaces of the film. It is important to note that this model predicts the excitation of only odd-integer half-wave resonances ($d = \lambda/2$, $3 \lambda / 2$, \dots) due to destructive interference at frequencies where the phonons are even-integer half-waves. Our data show clearly that both even and odd resonances are excited, however, which relates to the fact that the dynamic (and static) magnetization in these multilayers is certainly nonuniform. Furthermore, the boundary conditions likely differ at the bounding interfaces at the top and bottom of the multilayer since the interfaces themselves are different (Ta/Co on top and Co/Pd on bottom). The interior of the multilayer also promotes nonuniform magnetization due to the nonuniformity inherent in the proximity-induced magnetism in the Pd layers. The differences between our observations and the predictions of the model of \citet{Streib2018} underscore the importance of boundary conditions in the phonon pumping process, and it is probable that the complex magnetization depth profile associated with magnetic multilayers serves to enhance the phonon pumping.
\begin{figure}
  \centering
  \includegraphics{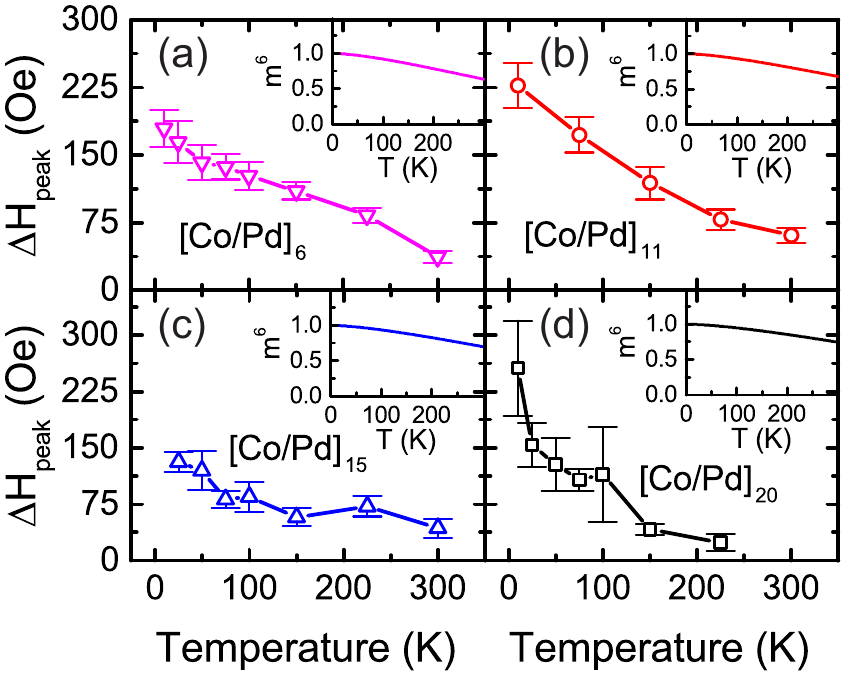}
  \caption{Temperature dependence of the peak full width at half maximum phonon pumping contribution to the FMR linewidth for the (a) [Co/Pd]$_6$, (b) [Co/Pd]$_{11}$, (c) [Co/Pd]${15}$, and (d) [Co/Pd]$_{20}$ multilayers. Insets show $m^6$ as a function of temperature.}\label{fig:tdep}
\end{figure}

The temperature dependence of the phonon pumping strength is expected to be primarily due to the dependence of magnetostriction on temperature \cite{Seavey1965,Streib2018}. It can be shown that since the magnetoelastic energy is quadratic in the magnetization cosines \cite{Kittel1958}, the magnetoelastic energy should scale with temperature as $m^3(T)$ \cite{Kittel1960,Callen1963a,Callen1966}, where $m(T) \equiv M(T)/M(0)$ is the reduced magnetization. (It is shown in the Supplemental Material that the interface anistropy energy exhibits $m^3$ scaling as expected \cite{Supplemental}.) As mentioned earlier, the phonon pumping amplitude depends on the square of the magnetoelastic energy and would therefore be expected to scale with temperature as $m^6$.

Figure \ref{fig:tdep} shows the temperature dependence of the peak phonon pumping contribution to the full width at half maximum (FWHM) FMR linewidths for all the multilayers. The damping enhancement depends quite strongly on temperature, with the effect being a factor of at least 4 greater at low temperature compared to room temperature. This depends on temperature much more strongly than $m^6$ (shown in the insets of Fig.\ \ref{fig:tdep}), which ranges from $\simeq$0.65 to 0.75 at 300~K in the four multilayers. It is unclear what causes the temperature dependence of the phonon pumping to be so strong, but there are several factors such as pinning and interlayer exchange coupling that also depend on temperature. We expect that the pinning of the dynamic magnetization at the exterior interfaces of the multilayer [Co/Ta on the top and Co/Pd on the bottom, see Fig.\ \ref{fig:fig1}(a)] becomes stronger at low temperature due to an increase in the interfacial anisotropy energy \cite{Callen1966}. In addition to the enhanced pinning, the interior Co/Pd interfaces will be affected by an increase in the interlayer exchange coupling at low temperatures that occurs for metallic nonmagnetic spacer layers \cite{Bruno1995,Postfach1994}, which may serve to enhance the phonon pumping.
\begin{figure}
  \centering
  \includegraphics{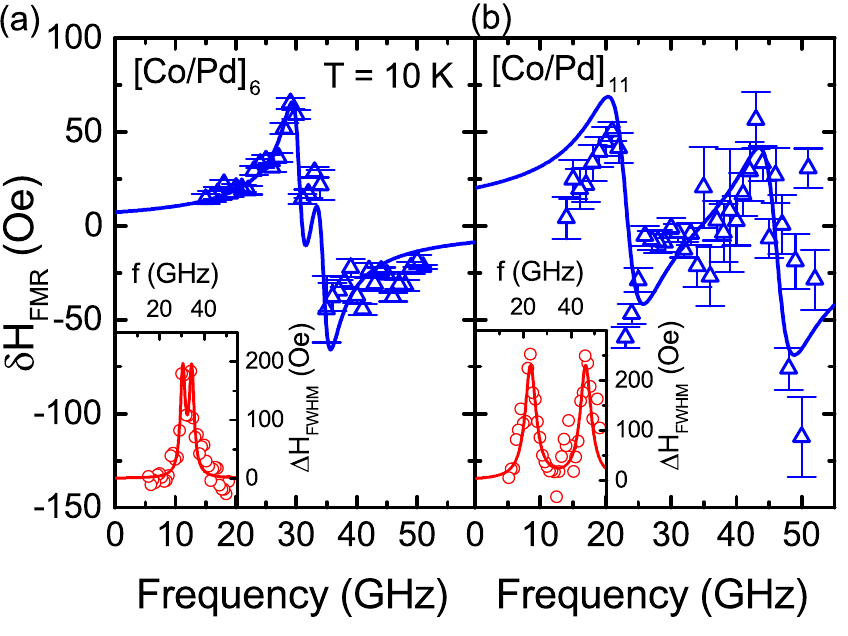}
  \caption{Shifts in the FMR field as a function of frequency at 10~K for the (a) [Co/Pd]$_6$ and (b) [Co/Pd]$_{11}$ multilayers. The insets in both panels show the corresponding linewidth enhancements as a function of frequency at 10~K. The solid curves in the main panels are predictions based on the fits of the linewidths in the insets using Kramers-Kronig relations.}\label{fig:fieldshifts}
\end{figure}

We now consider the effect of coupling to phonons on the effective field acting on the dynamic magnetization. Figure \ref{fig:fieldshifts} shows the observed shifts in FMR field as a function of frequency for the [Co/Pd]$_6$ and [Co/Pd]$_{11}$ multilayers at 10~K. The shifts are quantified by deviations of the FMR fields from the Kittel dispersion, so that the FMR field as a function of frequency is given by
\begin{equation}\label{eq:fieldshift}
  H_{FMR} = \omega / \gamma - H_{k,eff} + \delta H_{FMR}(\omega)~,
\end{equation}
where $H_{k,eff}$ is the uniaxial out-of-plane anisotropy field (containing both shape and interface contributions, defined here as positive for a PMA material), and $\delta H_{FMR}(\omega)$ is the frequency-dependent shift in FMR field due to phonon pumping. The Kramers-Kronig relations of linear response theory imply that an absorptive effect, here a resonant enhancement of the FMR linewidths, must be accompanied by a dispersive effect, i.e.\ a shift in the FMR field. Given that the absorptive response is  a Lorentzian [Eq.\ (\ref{eq:absorptive})], the field shifts caused by the dispersive response must be of the form
\begin{equation}\label{eq:dispersive}
  \delta H_{FMR} (\omega) = \sum_n -A_n \frac{\omega - n\omega_0}{(\omega-n\omega_0)^2 + (\delta \omega /2)^2}~,
\end{equation}
where the parameters $\delta \omega$, $\omega_0$, and $A$ are the same as in Eq.\ (\ref{eq:absorptive}).  The solid curves in the main panels of Fig.\ \ref{fig:fieldshifts} are predictions of the FMR field shifts based on the linewidth enhancements (the absorptive response): The parameters $\delta \omega$, $\omega_0$, and $A$ are determined from fits of the linewidths to Eq.\ (\ref{eq:absorptive}) [via Eq.\ (\ref{eq:pplinewidth})], and used to predict the FMR field shifts via Eq.\ \ref{eq:dispersive} with no free parameters. It can be seen from Fig.\ \ref{fig:fieldshifts} that there is good agreement between the observed and predicted FMR field shifts. We also note that the twin-peak structure seen in the linewidths of the [Co/Pd]$_6$ multilayer [inset of Fig.\ \ref{fig:fieldshifts}(a)] manifests as a kink between the two extrema in the FMR field shifts [main panel of Fig.\ \ref{fig:fieldshifts}(a)].

We conclude by emphasizing that the Co/Pd multilayer system is an ideal platform for phonon pumping due to the large magnetostriction and PMA, which opens the possibility of engineering devices that utilize this effect at zero applied field in the advantageous perpendicular configuration. As we have demonstrated, the frequency of the phonon pumping resonance is highly tunable by adjusting the number of Co/Pd repetitions---which notably does not significantly affect the magnitude of the PMA. It is therefore feasible to engineer a Co/Pd multilayer that experiences a phonon pumping resonance at zero external field.
\begin{acknowledgments}
This work was supported by SMART, a center funded by nCORE, a Semiconductor Research Corporation program sponsored by NIST. Parts of this work were carried out in the Characterization Facility, University of Minnesota, which receives partial support from NSF through the MRSEC program, and the Minnesota Nano Center, which is supported by NSF through the National Nano Coordinated Infrastructure Network, Award Number NNCI~-~1542202.
\end{acknowledgments}
\end{document}


\title{Supplemental Material for\\ ``Anomalous temperature dependence of phonon pumping by ferromagnetic resonance in [Co/Pd]$_n$ multilayers with perpendicular anisotropy''}
%
%
\maketitle
\tableofcontents
\section{SQUID magnetometry}\label{sec:SQUID}
\begin{figure}
  \centering
  \includegraphics{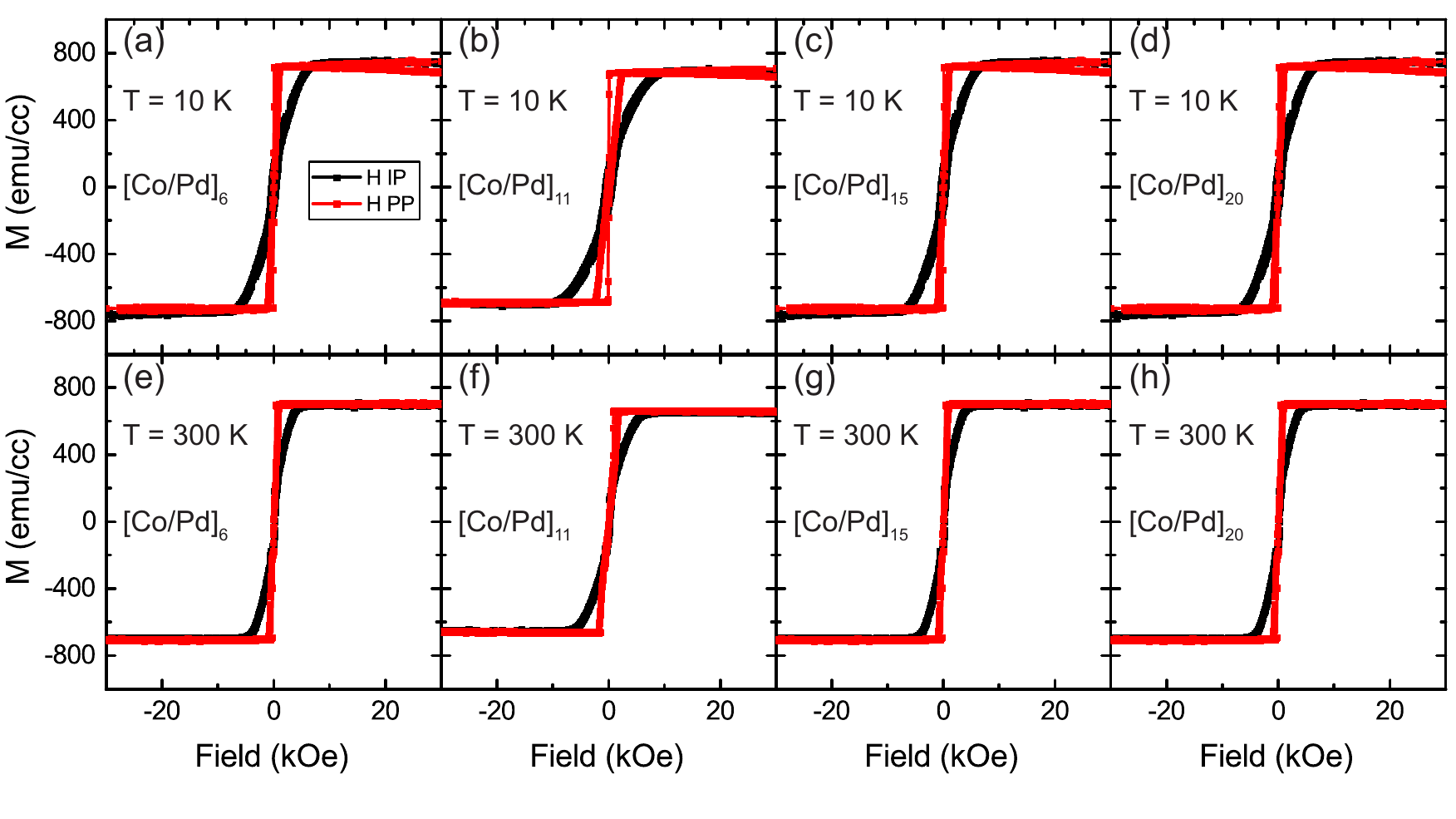}
  \caption{(a)--(d) $M$-$H$ loops with $T = 10$~K for field IP (black) and PP (red). (e)--(h) $M$-$H$ loops with $T = 300$~K for field IP (black) and PP (red).}\label{fig:MHloops}
\end{figure}

Superconducting quantum interference device (SQUID) magnetometry was performed on the [Co/Pd]$_n$ multilayers using a Quantum Design MPMS in the vibrating sample magnetometry (VSM) configuration. Figure \ref{fig:MHloops} shows $M$-$H$ loops for both field in-plane (black points) and perpendicular-to-plane (red points) at temperatures of 10~K (top row) and 300~K (bottom row) for the four multilayers. These $M$-$H$ loops confirm an out-of-plane easy axis in all of the multilayers.

\begin{figure}
  \centering
  \includegraphics{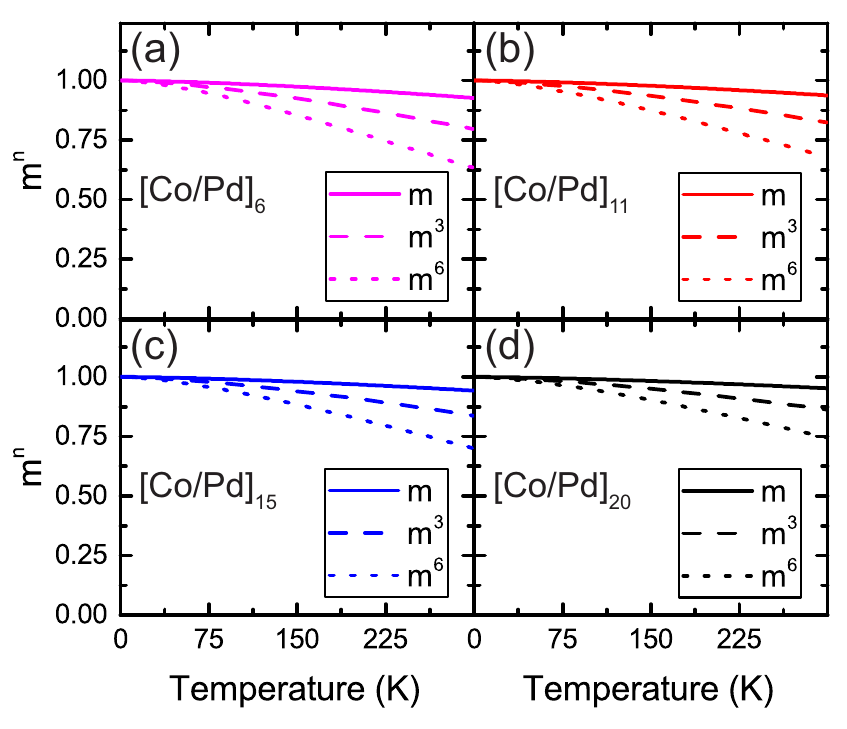}
  \caption{Reduced saturation magnetization $m \equiv M(T)/M(0)$ as a function of temperature raised to powers of 1 (solid line), 3 (dashed line), and 6 (dotted line) for the (a) [Co/Pd]$_6$, (b) [Co/Pd]$_{11}$, (c) [Co/Pd]$_{15}$, and (d) [Co/Pd]$_{20}$ multilayers}\label{fig:mscaling}
\end{figure}

SQUID was also used to measure the saturation magnetization as a function of temperature. The reduced saturation magnetization $m(T) \equiv M_s(T)/M_s(0)$ for the multilayers is shown in Fig.\ \ref{fig:mscaling}, along with the quantities $m^3$ and $m^6$. The saturation magnetization at a given temperature was determined by measuring the moment of the sample at four different fields (5, 7.5, 10, and 12.5~kOe) at which the sample was saturated for PP applied fields. The moment as a function of applied field was fit linearly, and the saturation moment was taken as the intercept of that fit since there is no diamagnetic or paramagnetic moment at zero applied field. The saturation magnetization was then determined by dividing the saturation moment by the volume of the multilayer.
\section{Absence of phonon pumping for in-plane magnetization}
\begin{figure}
  \centering
  \includegraphics{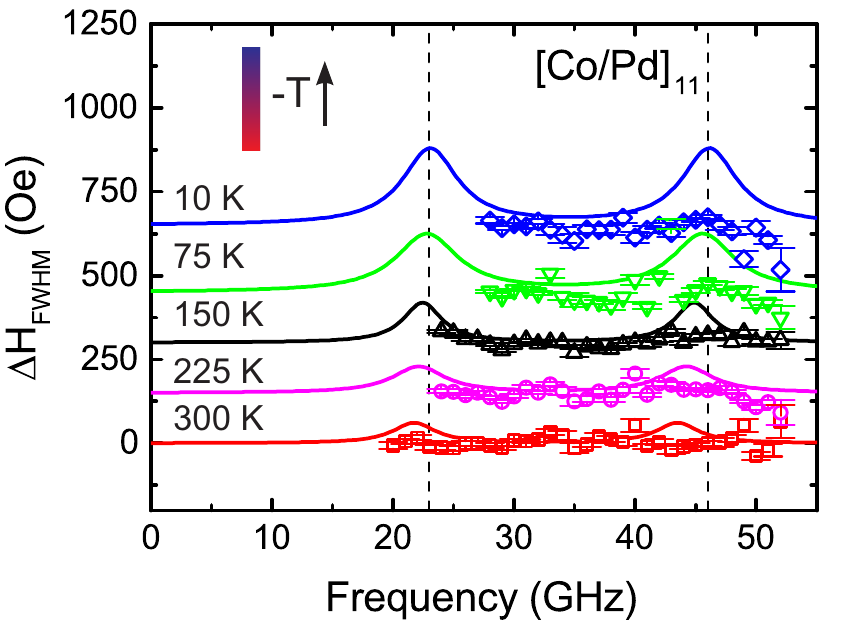}
  \caption{Ferromagnetic resonance linewidths---with Gilbert damping and inhomogeneous broadening contributions subtracted---as a function of frequency for the [Co/Pd]$_{11}$ multilayer with magnetization IP at temperatures of 10~K (blue diamonds), 75~K (green triangles), 150~K (black triangles), 225~K (magenta circles), and 300~K (red squares). The solid curves are fits to the corresponding FMR linewidths with PP magnetization. The data below 300~K were given a positive vertical offset so that the individual datasets could be more easily distinguished.}\label{fig:IPlinewidths}
\end{figure}
The phonon pumping effect is strongly mitigated for in-plane magnetization since the phonons driven by FMR propagate parallel to the magnetization (to leading order), and will therefore not propagate into the substrate \cite{Kittel1958,Streib2018,Sato2021}. We demonstrate this fact in Fig.\ \ref{fig:IPlinewidths}, where the FMR linewidths (with Gilbert damping and inhomogeneous broadening contributions subtracted) of the [Co/Pd]$_{11}$ multilayer with IP magnetization are shown at different temperatures. The lower frequency limit was set by the field below which the sample was unsaturated. The solid curves are the fits obtained from the PP linewidths. The disagreement between the IP FMR linewidths and the solid curves demonstrates the lack of thickness resonances, and therefore phonon pumping, for IP magnetization.
\section{Gilbert damping}
\begin{figure}
  \centering
  \includegraphics{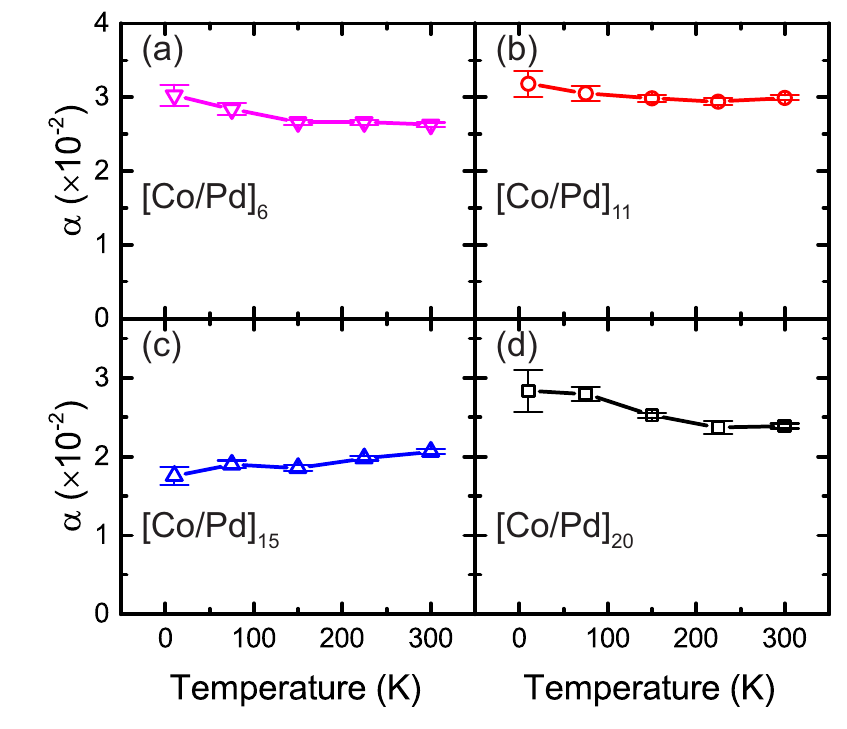}
  \caption{Gilbert damping as a function of temperature measured with applied field out of the plane for (a) [Co/Pd]$_6$, (b) [Co/Pd]$_{11}$, (c) [Co/Pd]$_{15}$, and (d) [Co/Pd]$_{20}$ multilayers.}\label{fig:alpha}
\end{figure}
The Gilbert damping $\alpha$ determines the linear dependence of the FMR linewidths. The FMR FWHM linewidths were used to determine the Gilbert damping through the relation
\begin{equation}\label{eq:pplinewidth}
  \Delta H_{FWHM} = \Delta H_0 + 2 \alpha \omega / \gamma + \Delta H_{ph}(\omega)~,
\end{equation}
where $\Delta H_0$ is the frequency-independent inhomogoneous broadening and $\Delta H_{ph}(\omega)$ is the phonon pumping contribution to the linewidth, the form of which is given in the main text. The Gilbert damping is shown as a function of temperature in Fig.\ \ref{fig:alpha} for the four multilayers. The Gilbert damping does not show a strong temperature dependence in any of the multilayers, showing a modest increase with temperature in the [Co/Pd]$_{15}$ multilayer and a modest decrease with temperature in the [Co/Pd]$_6$, [Co/Pd]$_{11}$, and [Co/Pd]$_{20}$ multilayers.
\section{Interface anisotropy}
\begin{figure}
  \centering
  \includegraphics{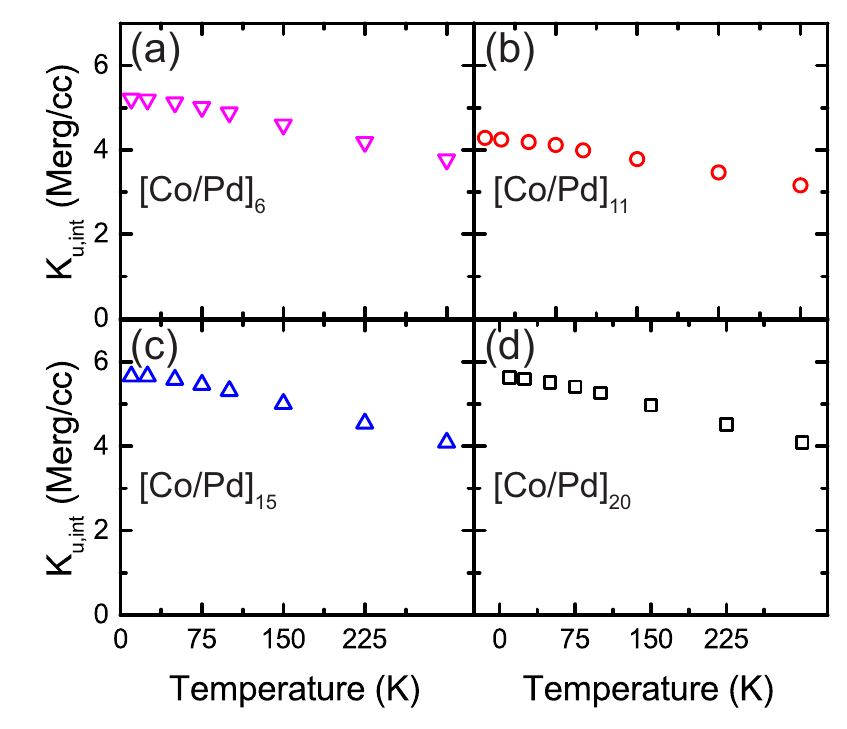}
  \caption{Uniaxial interface anisotropy $2 K_{u,int} / M_s = 4\pi M_s - H_{k,eff}$ as a function of temperature for the (a) [Co/Pd]$_6$, (b) [Co/Pd]$_{11}$, (c) [Co/Pd]$_{15}$, and (d) [Co/Pd]$_{20}$ multilayers.}\label{fig:kuintvsT}
\end{figure}
The interface anisotropy of the multilayers was determined from the relation $2 K_{u,int} / M_s = 4\pi M_s - H_{k,eff}$, where $H_{k,eff}$ is the net perpendicular anisotropy field measured with FMR and $M_s$ is the saturation magnetizion determined from SQUID VSM. The interface anisotropy $K_{u,int}$ is shown as a function of temperature in Fig.\ \ref{fig:kuintvsT} for the four multilayers.
\begin{figure}
  \centering
  \includegraphics{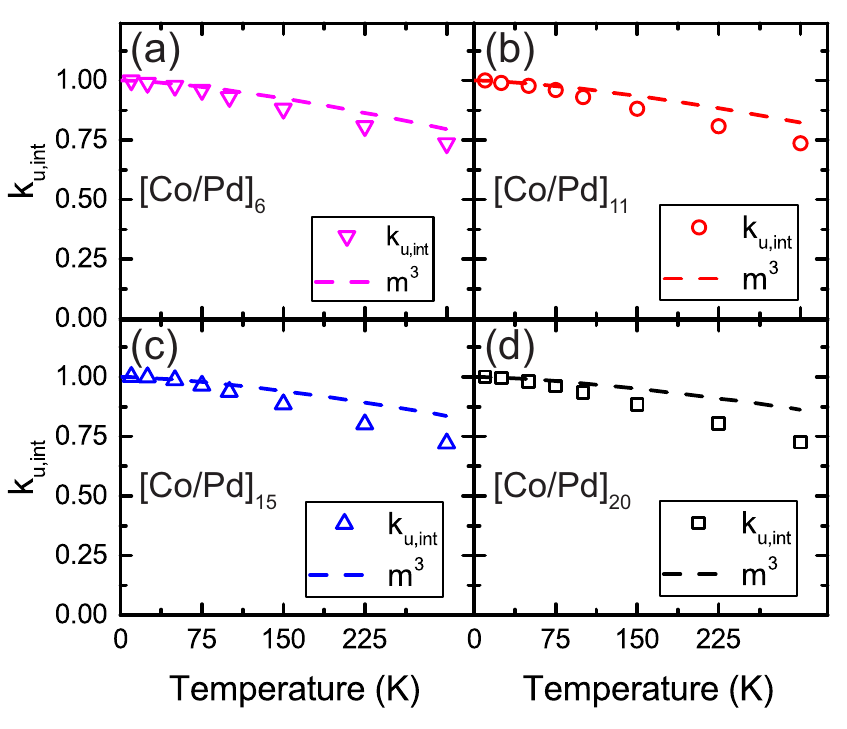}
  \caption{Reduced anisotropy field $k_{u,int} \equiv K_{u,int}(T)/K_{u,int}(10~\textrm{K})$ (symbols) and $m^3$ (dashed lines) as functions of temperature for the (a) [Co/Pd]$_6$, (b) [Co/Pd]$_{11}$, (c) [Co/Pd]$_{15}$, and (d) [Co/Pd]$_{20}$ multilayers.}\label{fig:kuintscaling}
\end{figure}

A uniaxial interface anisotropy is expected to scale with temperature as the cube of the saturation magnetization \cite{Callen1966}. The reduced interface anisotropy $k_{u,int} \equiv K_{u,int}(T)/K_{u,int}(10~\textrm{K})$ is shown as a function of temperature in Fig.\ \ref{fig:kuintscaling}, along with the quantity $m^3$. From Fig.\ \ref{fig:kuintscaling} it can be seen that the interface anisotropy shows good agreement with the $m^3$ scaling law.